\documentstyle[12pt,a4]{article}

\begin{document}

\newcommand{\nc}[2]{\newcommand{#1}{#2}}
\newcommand{\ncx}[3]{\newcommand{#1}[#2]{#3}}
\ncx{\pr}{1}{#1^{\prime}}
\nc{\pslsh}{\not{\! p}}
\nc{\nl}{\newline}
\nc{\np}{\newpage}
\nc{\nit}{\noindent}
\nc{\be}{\begin{equation}}
\nc{\ee}{\end{equation}}
\nc{\ba}{\begin{array}}
\nc{\ea}{\end{array}}
\nc{\bea}{\begin{eqnarray}}
\nc{\eea}{\end{eqnarray}}
\nc{\nb}{\nonumber}
\nc{\dsp}{\displaystyle}
\nc{\bit}{\bibitem}
\nc{\ct}{\cite}
\ncx{\dd}{2}{\frac{\partial #1}{\partial #2}}
\nc{\pl}{\partial}
\nc{\dg}{\dagger}
\nc{\cH}{{\cal H}}
\nc{\cL}{{\cal L}}
\nc{\cD}{{\cal D}}
\nc{\cF}{{\cal F}}
\nc{\cG}{{\cal G}}
\nc{\cJ}{{\cal J}}
\nc{\cQ}{{\cal Q}}
\nc{\tB}{\tilde{B}}
\nc{\tD}{\tilde{D}}
\nc{\tH}{\tilde{H}}
\nc{\tR}{\tilde{R}}
\nc{\tZ}{\tilde{Z}}
\nc{\tg}{\tilde{g}}
\nc{\tog}{\tilde{\og}}
\nc{\tGam}{\tilde{\Gam}}
\nc{\tPi}{\tilde{\Pi}}
\nc{\tcD}{\tilde{\cD}}
\nc{\tcQ}{\tilde{\cQ}}
\nc{\ag}{\alpha}
\nc{\bg}{\beta}
\nc{\gam}{\gamma}
\nc{\Gam}{\Gamma}
\nc{\bgm}{\bar{\gam}}
\nc{\del}{\delta}
\nc{\Del}{\Delta}
\nc{\eps}{\epsilon}
\nc{\ve}{\varepsilon}
\nc{\zg}{\zeta}
\nc{\th}{\theta}
\nc{\vt}{\vartheta}
\nc{\Th}{\Theta}
\nc{\kg}{\kappa}
\nc{\lb}{\lambda}
\nc{\Lb}{\Lambda}
\nc{\ps}{\psi}
\nc{\Ps}{\Psi}
\nc{\sg}{\sigma}
\nc{\spr}{\pr{\sg}}
\nc{\Sg}{\Sigma}
\nc{\rg}{\rho}
\nc{\fg}{\phi}
\nc{\Fg}{\Phi}
\nc{\vf}{\varphi}
\nc{\og}{\omega}
\nc{\Og}{\Omega}
\nc{\Kq}{\mbox{$K(\vec{q},t|\pr{\vec{q}\,},\pr{t})$ }}
\nc{\Kp}{\mbox{$K(\vec{q},t|\pr{\vec{p}\,},\pr{t})$ }}
\nc{\vq}{\mbox{$\vec{q}$}}
\nc{\qp}{\mbox{$\pr{\vec{q}\,}$}}
\nc{\vp}{\mbox{$\vec{p}$}}
\nc{\va}{\mbox{$\vec{a}$}}
\nc{\vb}{\mbox{$\vec{b}$}}
\nc{\Ztwo}{\mbox{\sf Z}_{2}}
\nc{\Tr}{\mbox{Tr}}
\nc{\lh}{\left(}
\nc{\rh}{\right)}
\nc{\ld}{\left.}
\nc{\rd}{\right.}
\nc{\nil}{\emptyset}
\nc{\bor}{\overline}
\nc{\ha}{\hat{a}}
\nc{\da}{\hat{a}^{\dg}}
\nc{\hb}{\hat{b}}
\nc{\db}{\hat{b}^{\dg}}
\nc{\hN}{\hat{N}}
\nc{\dle}{\del_{\epsilon}}
\nc{\dlk}{\del_{\kappa}}
\nc{\dla}{\del_{\alpha}}
\nc{\dlp}{\del^{\prime}}
\ncx{\abs}{1}{\left| #1 \right|}
\nc{\vs}{\vspace{2ex}}

\pagestyle{empty}

\begin{flushright}
NIKHEF/98-002\\
UTAS-PHYS-97-33 
\end{flushright}

\begin{center}
{\LARGE {\bf Off shell $\kg$-symmetry of the superparticle and the
spinning superparticle}} \\

\vspace{4ex}

{\large P.D.\ Jarvis$^*$ and J.W.\ van Holten} \\
NIKHEF, Amsterdam NL \\
\vspace{4ex}

{\large J.\ Kowalski-Glikman} \\
ITP, Univ.\ of Wroclaw PL

\vspace{5ex}
\today \\
\vspace{5ex}

{\bf Abstract} \\
\end{center}
\vspace{2ex}

\nit
{\small
The spinorial local world-line $\kg$-symmetry of the covariant Brink-Schwarz
formulation of the 4-$D$ superparticle is abelian in an off-shell phase-space
formulation. The result is shown to generalize to the extended spinorial 
transformations of the spinning superparticle.
}

\nit
{\small}
\vfill
\nit
{\footnotesize $^*$ 
Dept of Physics, University of Tasmania, Hobart Tas 7001 Australia} 

\np

\pagestyle{plain}
\pagenumbering{arabic}

\section{Introduction}

The space-time covariant formulation of super $p$-branes is known
to have a local fermionic invariance on the world manifold, first
discovered by Siegel for the superparticle \ct{siegel}, and
subsequently generalized to the case of superstrings in \ct{GS}.
This invariance helps to balance the number of commuting and
anti-commuting degrees of freedom in models with the boson and
fermion variables naturally belonging to different representations
of the Lorentz group of the target space-time. Indeed, the parameter 
of this transformation is an anti-commuting space-time spinor $\kg$, 
varying in an arbitrary way over the world manifold. In this sense
the $\kg$-invariance is a supersymmetry. However, it is not a 
classic supersymmetry as provided by the grading of space-time 
algebras, like the Poincar\'{e} or conformal superalgebras. Although 
it can be related to space-time supersymmetry by gauge-fixing in the 
light-cone formulation, the algebraic structure underlying the
$\kg$-symmetry has not been fully elucidated. Rather, in most 
instances it has been found as a local symmetry of specific actions, 
and not by implementing a given algebraic structure in the context 
of the general principles of field theory. A related problem is, 
that in most examples the symmetry has only been realized on-shell, 
i.e.\ modulo the equations of motion of the specific model considered. 
A step forward is therefore provided by examples of $\kg$-invariance 
which are realized off-shell, i.e.\ with an algebraic structure 
independent of the equations of motion.

In this paper we consider the example of the superparticle and its   
generalization with additional local world-line supersymmetry, known 
as the spinning superparticle \ct{aoy}-\ct{bergh}. Both models are 
reparametrization invariant on the world-line, and posses a 
proper-time dependent $\kg$-invariance. In addition, the spinning 
superparticle possesses a local world-line supersymmetry, bringing 
along a bosonic counterpart of $\kg$-symmetry with a {\em commuting} 
spinor parameter $\ag$ \ct{aoy,jw1}. 

We present results showing that the first order (phase-space) 
formulations of these models have some remarkable features: they 
contain a $(D = 1)$ gauge field for local $\kg$-symmetry \ct{jkg2}, 
the algebra of $\kg$-transformations and other local world-line 
symmetries closes off-shell, and the structure of the algebra 
becomes abelian. This is in strong contrast with the on-shell 
results \ct{bergh,huq}. As the $\kg$-transformations also involve the 
momenta explicitly, we conclude that the phase-space formulation 
of these models seems the more natural one, and we make some
remarks about prospects for quantisation. How our present formulation 
is to be implemented for higher-dimensional objects like strings 
or membranes remains to be investigated. 

\section{Phase-space description of the superparticle} 

The standard action for a free massless superparticle described by 
bosonic and fermionic coordinates $(x^{\mu}, \th_{\ag})$ in 
4-$D$ space-time is 

\be
S_{conf}\, =\, \int d\tau\, \frac{1}{2e}\, \lh \dot{x}^{\mu} -  
\bar{\th} 
   \gam^{\mu} \dot{\th} \rh^2.
\label{1.1}
\ee

\nit
Here a dot denotes a proper-time derivative, and $e$ is the einbein 
variable making the action reparametrization invariant: under 
an arbitrary transformation of the affine parameter (proper time)
$\tau \rightarrow \tau^{\prime}$ the action (\ref{1.1}) is invariant 
provided 

\be
x^{\prime\, \mu}(\tau^{\prime}) = x^{\mu}(\tau), \hspace{2em} 
\th_{\ag}^{\prime}(\tau^{\prime})\, =\, \th_{\ag}(\tau), \hspace{2em} 
e^{\prime}(\tau^{\prime})\, =\, \frac{d\tau}{d\tau^{\prime}}\, e(\tau).
\label{1.2}
\ee

\nit 
For ease of notation, in the following we regularly suppress indices on 
$\th$ and other spinor variables.  

The momenta conjugate to the coordinates $(x^{\mu}, \th_{\ag})$ are 

\be
p^{\mu} = \frac{1}{e}\, \lh \dot{x}^{\mu} - \bar{\th} \gam^{\mu} 
 \dot{\th} \rh, \hspace{2em} 
\pi = - \pslsh \th.
\label{1.3}
\ee

\nit
The first-class constraint imposed by reparametrization invariance 
is that the momenta are light-like:

\be
p_{\mu}^2 \, =\, 0,
\label{1.4}
\ee

\nit
showing that the particle is massless indeed. The equations of motion 
imply the conservation of momentum:

\be
\dot{p}_{\mu}\, =\, 0,
\label{1.4.1}
\ee

\nit
as required by translation invariance in the target space-time.
The remaining equations of motion for the fermionic coordinates are  

\be
\dot{\pi}\, =\, \pslsh \dot{\th},
\label{1.5}
\ee

\nit
which implies

\be
\pslsh \dot{\th}\, =\, 0.
\label{1.6}
\ee

\nit
As noted by Siegel \ct{siegel}, the first-class constraint (\ref{1.4}) 
and momentum conservation imply that this equation has zero-modes of the 
form

\be
\th\, =\, \pslsh \kg,
\label{1.7}
\ee

\nit
where $\kg(\tau)$ is an arbitrary spinor-valued function of $\tau$. 
The local $\kg$-invariance found by Siegel generates precisely these 
zero-modes.

We now present a phase-space action for the superparticle in which the 
coordinates and momenta are independent variables, and which reduces 
to the above model in terms of solutions of the equations of motion 
\ct{jkg2}. This model has the advantage that the structure of the 
$\kg$-symmetry simplifies considerably.

The starting point for this phase-space description of the superparticle 
is the action

\be
S_{phase}\, =\, \int d\tau\, \lh - \frac{e}{2}\, p_{\mu}^2 +
  p_{\mu} \lh \dot{x}^{\mu} + \bar{\fg} \gam^{\mu} \th \rh
  + \bar{\pi} \lh \fg - \dot{\th} \rh \rh,
\label{1.8}
\ee

\nit
where $(x^{\mu}, p_{\mu})$ are space-time vectors, and $(\th, \fg, \pi)$ 
space-time Majorana spinors. The equations of motion for the new variables  
guarantee
that $(p_{\mu}, \pi)$ are given by the expressions (\ref{1.3}), and 
furthermore that on shell

\be
\fg\, =\, \dot{\th}.
\label{1.9}
\ee

\nit
As a result the action $S_{phase}$ indeed reproduces the results of 
the configuration space formulation based on $S_{conf}$. Like the
original action, $S_{phase}$ is reparametrization invariant, provided 
$(x^{\mu}, p_{\mu}; \th_{\ag}, \pi_{\ag})$ transform as world-line  
scalars,
and $(e,\fg_{\ag})$ like bosonic, respectively fermionic, world-line 
vectors. In particular the transformation rule for the anti-commuting 
variables $\fg_{\ag}$ is analogous to that of the einbein $e$:

\be
\fg_{\ag}^{\prime}(\tau^{\prime})\, =\, \frac{d\tau}{d\tau^{\prime}}\, 
 \fg_{\ag}(\tau).
\label{1.10}
\ee

\nit
There is a natural explanation for this transformation character of the 
spinor $\fg$. This becomes clear from the local $\kg$-symmetry  
which plays
the same role here as previously: to reduce the number of physical  
fermion
degrees of freedom in the solutions for $\th$ by generating zero-modes. 
The infinitesimal transformations with spinor parameter  
$\kg_{\ag}(\tau)$,
and the infinitesimal reparametrizations with scalar parameter  
$\xi(\tau)
= \tau - \tau^{\prime}$, which leave the action $S_{phase}$ invariant 
modulo boundary terms read

\be
\ba{ll}
\dsp{ \del e = -2 \bar{\th} \dot{\kg} + 2  \bar{\fg} \kg +
  \frac{d(\xi e)}{d\tau}, } & \dsp{
  \del \fg = \pslsh \dot{\kg} + \frac{d(\xi \fg)}{d\tau}, }\\ 
  & \\
\dsp{ \del x^{\mu} = \bar{\kg} \gam^{\mu} \pi +
  \xi\, \frac{dx^{\mu}}{d\tau}, } & \dsp{
  \del \th = \pslsh \kg + \xi\, \frac{d \th}{d\tau}, }\\
  & \\
\dsp{ \del p_{\mu} = \xi\, \frac{dp_{\mu}}{d\tau}, } & \dsp{
  \del \pi = \xi\, \frac{d \pi}{d\tau}. }
\ea
\label{1.11}
\ee

\nit
Clearly, the $\kg$-transformation of $\th$ generates the fermion
zero-modes, whilst that of $\fg$ generates the proper-time derivative 
of these modes. Thus $\fg$ acts as a $D = 1$ gauge field for local
$\kg$-transformations, and it is not surprising that it transforms
as a world-line vector under reparametrizations, rather than as a
scalar.

The commutator algebra of these infinitesimal transformations has a 
very simple structure. Computing the commutator of two transformations 
with parameters $(\kg_1, \xi_1)$ and $(\kg_2, \xi_2)$ results in a
similar transformation

\be
\left[ \del(\kg_2, \xi_2), \del(\kg_1, \xi_1) \right]\, =\,
  \del(\kg_3, \xi_3),
\label{1.12}
\ee

\nit
with the parameters on the right-hand side given by

\be
\kg_3 = \xi_1 \dot{\kg}_2 - \xi_2 \dot{\kg}_1, \hspace{2em}
 \xi_3 = \xi_1 \dot{\xi}_2 - \xi_2 \dot{\xi}_1.
\label{1.13}
\ee

\nit
We observe, that the $\kg$-transformations commute among themselves, 
and the parameter $\kg$ transforms as a scalar under world-line
reparametrizations. As these results do not require the use of any
equations of motion, this algebra is closed in a model-independent way; 
indeed, it is shown below that the abelian nature of the
$\kg$-transformations survives in a more complicated model including 
additional physical degrees of freedom, obtained as the world-line
supersymmetric extension of the superparticle.

On the other hand, it is presently not clear to us whether such
a result would survive in a theory with interactions, which might
generate a non-linear extension of the $\kg$-transformations;
this is what happens in general relativity, where the general
coordinate transformations can be interpreted as a non-linear
extension of the abelian gauge transformations of a free massless
spin-2 tensor field. It is also true that in different
Lagrangians with equivalent Hamiltonian
descriptions, various gauge symmetries can be
presented in different ways (as between the
first order formulation of the scalar relativistic particle with
explicit einbein as gauge field, versus
the second-order form with Lagrange multiplier to carry the 
diffeomorphism symmetry).

\section{The spinning superparticle}

The spinning superparticle is a superparticle model with both
rigid target space-time and local world-line supersymmetry
\ct{aoy}-\ct{bergh}. To construct actions with local world-line
supersymmetry we use the conventions of \ct{jw1}, which also
describes the second-order action (formulated in configuration space) 
for the free $D = 4$ spinning superparticle in our notation, both in 
superfield and component form.

In this section we construct a first-order (phase-space) action,
equivalent to the standard second-order action after elimination of 
momenta and lagrange multipliers. The construction requires two sets 
of super multiplets $(\Sg^{\mu}, \Fg_{\ag})$ for the bosonic and
fermionic particle coordinates $(x^{\mu}, \th_{\ag})$, transforming 
respectively as a vector and a spinor under the target-space Lorentz 
group; in addition there are two multiplets $(\Og_{\mu}, \Pi_{\ag})$ 
containing their conjugate momenta $(p_{\mu}, \pi_{\ag})$. 
Finally there are gauge-multiplets $(E,Y_{\ag})$ for local world-line 
supersymmetry and local world-line $\kg$-symmetry.

The component content of these superfields is

\be
\ba{ll}
\Sg^{\mu} = (x^{\mu}, \psi^{\mu}), & \Fg = (\th, h), \\
 & \\
\Og_{\mu} = (\og_{\mu}, p_{\mu}), & \Pi = (\pi, n), \\
 & \\
E = (e, \chi), & Y = (y, \eta).
\ea
\label{2.1}
\ee

\nit
As before we have suppressed indices of spinor components on
$(\Fg,\Pi,Y)$. The superfields $(\Sg^{\mu}, Y_{\ag}, E)$ have a
commuting world-line scalar $(x^{\mu}, y_{\ag})$ or vector $(e)$
as their first component, whilst the first components of $(\Fg_{\ag}, 
\Og_{\mu}, \Pi_{\ag})$ are anti-commuting world-line fermions.

The supersymmetry transformation rules and the construction of
invariant actions are discussed in \ct{jw1}. For the supersymmetry
gauge variables $(e, \chi)$ the infinitesimal supertransformations are 

\be
\dle\, e = -2i\ve \chi, \hspace{3em} \dle\, \chi = \dot{\ve}.
\label{2.2}
\ee

\nit
For scalar multiplets like $Y = (y, \eta)$ they take the
form

\be
\dle\, y = - i \ve \eta, \hspace{3em}
\dle\, \eta = \ve \frac{1}{e}\, \cD_{\tau} y = \ve \frac{1}{e}\,
 \lh \dot{y} + i \chi \eta \rh.
\label{2.3}
\ee

\nit
Finally, for the fermionic multiplets such as $\Fg = (\th, h)$ the
infinitesimal component transformations are

\be
\dle\, \th = \ve h, \hspace{3em} \dle\, h = - i \ve \frac{1}{e}\,
 \cD_{\tau} \th = - i \ve \frac{1}{e}\, \lh \dot{\th} - \chi h \rh.
\label{2.4}
\ee

\nit
The component action is constructed from the following fermionic
superfield expression

\be
\Lb\, \equiv\, (\lb, \ell )\,
  =\, \lh -\frac{1}{2}\, \cD \Og^{\mu}\, +\, \cD^2 \Sg^{\mu} \rh\,
  \times \Og_{\mu}\, +\, i \cD \Og_{\mu} \times
  \lh \overline{Y} \gam^{\mu} \Fg \rh\, -\,
  i \lh \overline{\cD \Fg} - \overline{Y} \rh \times \Pi,
\label{2.5}
\ee

\nit
where $\cD$ denotes the super derivative \ct{jw1}. The explicit result 
for the locally supersymmetric component action is

\be
\ba{lll}
S & = & \dsp{ \int d\tau \lh e \ell - i \chi \lb \rh }\\
 & & \\
 & = & \dsp{ \int d\tau\, \lh -\frac{e}{2}\, p_{\mu}^2 + i
  \lh \ps^{\mu} + i \bar{y} \gam^{\mu} \th - \frac{1}{2}\,  
\og_{\mu} \rh
  \dot{\og}^{\mu} - i e \bar{n} \lh y - h\rh \rd }\\
 & & \\
 & & \dsp{ ~~ \hspace{2em} \ld +\, p_{\mu}\, \lh \cD_{\tau} x^{\mu} 
     + e \bar{\eta} \gam^{\mu} \th + i e \bar{y} \gam^{\mu} h \rh
     - \bar{\pi} \lh \dot{\th} - e \eta
     - \chi y \rh \rh.}
\ea
\label{2.6}
\ee

\nit
In addition to local world-line supersymmetry, the action is invariant 
under local $\kg$-transformations which are a direct extension of those 
for the ordinary superparticle:

\be 
\ba{ll}
\dlk e = -2 \bar{\th} \dot{\kg} + 2 e \bar{\eta} \kg + 2 \chi  
\bar{y} \kg,
  & \dlk \chi = 0, \\
  & \\
\dlk x^{\mu} = \bar{\kg} \gam^{\mu} \pi, & \dsp{ \dlk \ps^{\mu} =
 - i \bar{y} \gam^{\mu} \pslsh \kg + i \frac{\dlk e}{2e}\,
 \bar{y} \gam^{\mu} \th, } \\
 & \\
\dlk \th = \pslsh \kg, & \dsp{ \dlk h = - \frac{\dlk e}{2e}\,  
h, }\\
 & \\
\dlk \og_{\mu} = 0, &  \dlk p_{\mu} = 0, \\
 & \\
\dlk \pi = 0, & \dsp{ \dlk n = - \frac{\dlk e}{2e}\, n, }\\
 & \\
\dsp{ \dlk y = - \frac{\del e}{2e}\, y, } &  \dsp{ \dlk (e\eta) =
     \pslsh \dot{\kg} + \frac{\dlk e}{2e}\, \chi y. }
\ea
\label{2.7}
\ee

\nit
Working in $D = 4$ space-time dimensions, the parameter $\kg(\tau)$ 
is an anti-commuting Majorana spinor. Comparing the action and 
$\kg$-transformations to those of the standard superparticle, it is 
convenient to introduce new variables

\be
\ba{ll}
\fg \equiv e \eta + \chi y, & \\
 & \\
\lb^{\mu} \equiv \ps^{\mu} + i \bar{y} \gam^{\mu} \th, &
 \zg^{\mu} \equiv \og^{\mu} - \lb^{\mu}.
\ea
\label{2.7.2}
\ee

\nit
The variable $\fg$ acts as the gauge field of $\kg$-transformations on 
the world line, whilst the anti-commuting Lorentz vectors $\lb^{\mu}$ 
and $\zg^{\mu}$ are $\kg$-invariant:

\be
\ba{ll}
\dlk \fg = \del \lh e \eta + \chi y \rh = \pslsh \dot{\kg}, &
 \dlk e = -2\bar{\th} \dot{\kg} + 2 \bar{\fg} \kg, \\
 & \\
\dlk \lb^{\mu} = 0, & \dlk \zg^{\mu} = 0.
\ea
\label{2.7.1}
\ee

\nit
In addition, the supersymmetric structure of the model allows yet
another local world-line invariance with a {\em commuting} Majorana 
spinor parameter $\ag$ \ct{aoy,jw1}, which is implemented here by \nl 

\be
\ba{ll}
\dla e = -2ie \bar{\ag} \lh y + h \rh, & \dla \chi = 0, \\
 & \\
\dsp{ \dla y = \pslsh \ag - \frac{\dla e}{2e}\, y, } &
 \dla \fg = \dla (e\eta + \chi y) = 0, \\
 & \\
\dla x^{\mu} = 0, & \dla \lb^{\mu} = \dla \lh \ps^{\mu} +
 i \bar{y} \gam^{\mu} \th \rh = 0, \\
 & \\
\dla \th = 0, &
     \dsp{ \dla h = \pslsh \ag - \frac{\dla e}{2e}\, h, }\\
\ea
\label{2.8}
\ee
 
~~~~~~~~~~~~~~~~~~~~\begin{tabular}{ll}
$\dla \zg_{\mu} = 0,$ & $\dla p_{\mu} = 0, $ \\
 & \\
$\dla \pi = 0,$ & $\dsp{ \dla n = - \frac{\dla e}{2e}\, n.}$  
\end{tabular} \hfill (\ref{2.8}, cont'd)
\vs 

\nit
From this  point of view the simplest form of the
$\ag$- and $\kg$-transformation rules is obtained by defining

\be
u = \frac{\sqrt{e}}{2}\, (y - h), \hspace{2em}
v = \frac{\sqrt{e}}{2}\, (y + h), \hspace{2em}
w = 2 \sqrt{e}\, n,
\label{2.9}
\ee

\nit 
whilst at the same time redefining the spinor parameter by 

\be
\ag^{\prime}\, =\, \sqrt{e}\, \ag. 
\label{2.9.1}
\ee

\nit 
This is allowed, because $\ag(\tau)$ is an arbitrary function 
of proper time. Then, dropping the prime on the parameter: 
$\ag^{\prime} \rightarrow \ag$, the complete spinor-symmetry 
transformation rules become

\be
\ba{ll}
\dlp e = 2 \lh \bar{\fg} \kg - \bar{\th} \dot{\kg}\rh
     - 4 i \bar{\ag} v, & \dlp v = \pslsh \ag, \\
 & \\
 \dlp \fg = \pslsh \dot{\kg}, & \\
 & \\
\dlp \th = \pslsh \kg, &
\dlp x^{\mu} = \bar{\kg} \gam^{\mu} \pi,
\ea
\label{2.10}
\ee

\nit
with all other variables invariant: $\dlp
(\chi,\lb^{\mu},\zg^{\mu},p_{\mu},\pi,u,w) = 0$. Comparison of the 
two representations of $\kg$-symmetry in (\ref{1.11}) for the 
scalar superparticle and (\ref{2.10}) for the spinning superparticle 
are identical.

In terms of the new variables the action becomes

\be
\ba{lll}
S & = & \dsp{ \int d\tau\, \lh -\frac{e}{2}\, p_{\mu}^2\, +\,
  p_{\mu} \lh \dot{x}^{\mu} + i \chi \lb^{\mu} +
  \bar{\fg} \gam^{\mu} \th + i \bar{v} \gam^{\mu} v -
  i \bar{u} \gam^{\mu} u \rh \rd } \\
  & & \\
  & & \dsp{ \ld ~~ \hspace{2em}
  -\, \frac{i}{2}\, \zg_{\mu} \dot{\zg}^{\mu}\,
  +\, \frac{i}{2}\, \lb_{\mu} \dot{\lb}^{\mu}\,
  -\, i \bar{w} u\, +\, \bar{\pi} \lh \fg -\dot{\th}\rh \rh. }
\ea
\label{2.7.4}
\ee

\nit
Of course, the above variable redefinitions do not respect the multiplet 
structure of world-line supersymmetry, and checking invariance of the 
new action under world-line supersymmetry and reparametrizations is more 
complicated.

Turning to the algebra of infinitesimal spinorial transformations
(\ref{2.10}), it is readily observed that they all commute:

\be
\left[ \dlp(\kg_2, \ag_2), \dlp(\kg_1,\ag_1) \right]\, =\, 0.
\label{2.11}
\ee

\nit
For the $\kg$-transformations this follows directly from the
commutator algebra (\ref{1.12}); for the $\ag$-transformations
it follows because commuting Majorana spinors $\ag_i$ satisfy the
transposition rule $\bar{\ag}_1 \gam^{\mu} \ag_2 = \bar{\ag}_2
\gam^{\mu} \ag_1$. Thus the extended algebra of off-shell
$\kg$- and $\ag$-transformations is again abelian for free
spinning superparticles. 

Although the commutator algebra (\ref{2.11}) has been established 
for the transformations with parameter $\ag$ redefined as in 
(\ref{2.9.1}), the abelian character holds for the original version 
(\ref{2.8}) as well. This is because in the original version the 
commutator of two $\ag$-transformations is only modified by terms 
containing the commutator on the einbein: 

\be
\left[ \dlp_{\ag_2}, \dlp_{\ag_1} \right]\, \propto 
 \frac{1}{\sqrt{e}}\, \left[ \del_{\ag_2}, \del_{\ag_1} \right]\,
 \sqrt{e},
\label{2.12}
\ee

\nit
which vanishes. This we have checked by an explicit calculation. 
However, the commutator of $\ag$ and $\kg$ transformations is 
modified in a non-trivial way, which can only be redressed by 
redefining the anti-commuting spinor parameters $\kg$ as well. 

In summary, we have found a phase-space version of spinorial 
$\kg$ tranformations for the superparticle, and a bosonically 
extended version for the spinning superparticle, with an 
off-shell abelian commutator algebra. The basic multiplets  
of local $\kg$- and $\ag$-transformations are $(e,\fg,\th_{\ag},
p_{\mu})$ and $(x^{\mu},\pi)$ for $\kg$ transformations, and
$(e,v)$ for the $\ag$-transformations. The transformation rules
are given in (\ref{2.10}), the corresponding action in (\ref{2.7.4}).

Finally, we make some remarks on prospects for quantisation
from the perspective of the present work. Although systems with 
reducible gauge symmetries are notoriously intractible (see for 
example \cite{kallosh} for a recent discussion), our present 
off-shell phase space formulation may allow new approaches to
covariant gauge fixing. For example for the simple (scalar)
superparticle introduce anti-commuting scalar ghosts
$(b,c)$ and lagrange multiplier $k$ for reparametrizations,
and commuting spinor ghosts $(\beta, \gamma)$ with lagrange
multiplier $\rho$ for $\kappa$-symmetry. The anti-commuting
nilpotent variational BRST derivative $\delta$ then is,
for the physical variables, formally given by (\ref{1.11}) above
with ghosts $c, \gamma$ in place of the parameters $\xi,
\kappa$ of the infinitesimal diffeomorphisms and Siegel
transformations respectively. For the ghost variables the 
non-zero BRST variations read:
\[ \delta c = c \dot{c}, \hspace{1em} \delta \gamma = c \dot{\gamma},
    \hspace{1em} \delta b = k, \hspace{1em} \delta \beta = \rho, \]  
for which nilpotence may be easily verified. If the
action of diffeomorphisms and $\kappa$-supersymmetries on phase space 
$(x^\mu, p_\mu, \theta, \pi)$ is {\it free} for generic 
points\footnote{On shell with $\kappa =\pslsh \kappa'$, we have 
however $\dlk \theta = \dlk \pi = \dlk p = 0$ but 
$\dlk x^\mu  \propto p^\mu \bar{\kappa}'\pi$}, 
a suitable gauge fixing in the total space $(x^\mu, p_\mu, \theta, 
\pi, e, \phi)$ is induced \cite{govaerts} from that on the space 
of connections $(e, \phi)$. In these circumstances gauge classes 
such as $\dot{\phi}= \dot{e} = \mbox{const}$ may be admissible.
Such conditions are natural concomitants of the full BRST-BFV 
formalism, in which connections with the representation theory 
of extended spacetime superalgebras have recently been reported 
\cite{ospwigws}. Further work along these lines is in progress.
\vs 

\nit
{\bf Acknowledgement:}\\
PDJ thanks the theory group at NIKHEF for hospitality during a visit
on leave from the University Tasmania.

\end{document}